\documentclass[aps,showpacs,11pt]{revtex4}
\usepackage{dcolumn}
\usepackage{graphicx}
\usepackage{amsmath}
\usepackage{amsfonts}
\usepackage{amssymb}
\usepackage{psfrag}
\usepackage{wrapfig}
\usepackage{subfigure}
\usepackage{makeidx}
\usepackage{bm}
\usepackage{epsf}
\usepackage{hyperref}
\usepackage{color}
\usepackage{multirow,dcolumn}
\usepackage{graphicx}

\begin{document}

\title{Gravitational collapse of massless scalar field in $f(R)$ gravity}

\author{Cheng-Yong Zhang}
\email{zhangcy0710@pku.edu.cn}
\affiliation{Center for High-Energy Physics, Peking University, Beijing
	100871, China}

\author{Zi-Yu Tang}
\email{tangziyu@sjtu.edu.cn}
\affiliation{Center of Astronomy and Astrophysics, Department of Physics
	and Astronomy, Shanghai Jiao Tong University, Shanghai 200240, China}

\author{Bin Wang}
\email{wang\_b@sjtu.edu.cn}
\affiliation{Center of Astronomy and Astrophysics, Department of Physics
	and Astronomy, Shanghai Jiao Tong University, Shanghai 200240, China}


\begin{abstract}
We study the spherically symmetric gravitational collapse of massless scalar matter field in
asymptotic flat spacetime in  the Starobinsky $R^2$ gravity, one specific model in the $f(R)$ gravity. In the Einstein frame of $f(R)$ gravity, an additional scalar field arises due to the
conformal transformation. We find that besides the usual competition between gravitational
energy and kinetic energy in the process of gravitational collapse,
the new scalar field brought by the conformal transformation adds one more competing force in the dynamical system. The dynamical competition can be controlled by tuning the amplitudes of the initial perturbations of the new scalar field and the matter field.  To understand the physical reasons behind these phenomena, we analyze the gravitational potential behavior and calculate the Ricci scalar at center with the change of initial amplitudes  of perturbations. We find rich physics on the formation of black holes through gravitational collapse in $f(R)$ gravity.
\end{abstract}


\maketitle

\section{Introduction}
The dynamical evolution of an isolated system in general relativity is governed by the Einstein-matter equations and it can end up in qualitatively different stable end states, either with the formation of a single black hole in collapse or completely dispersion of the mass-energy to infinity. For a massless scalar field in spherical  symmetry, these are the only possible end states. The fate of the evolution of an isolated system in general relativity is determined by the dynamical competition which  can be controlled by tuning  parameters in the initial conditions \cite{Christodoulou_87, Christodoulou_91, Christodoulou_94, Wald_97}. If the parameter $p$ is less than some critical value $p_\star$, the kinetic energy of the field will completely disperse the mass-energy of the system to infinity; while if $p>p_\star$, the gravitational potential  dominates and a black hole forms. It is intriguing that the phenomenology of existing a threshold of black hole formation becomes apparent and universal in a large class of collapse models in general relativity. This phenomenology, which includes scaling, self-similarity and universality, is analogous to the critical behavior in statistical mechanics. Critical phenomena in gravitational collapse in Einstein gravity were first discovered by Choptuik in the model of a spherical symmetric, massless scalar field minimally coupled to general relativity \cite{Choptuik_92,Choptuik_93,Choptuik_94}. Important results of gravitational collapse in the best-studied models  have been reviewed in spherical symmetry in asymptotically flat spacetimes \cite{Choptuik_98,WangAn_01,Gundlach_07}. The gravitational collapse in de Sitter Einstein gravity has also been investigated in  \cite{Hod_97,Iwashita_05,Zhang_15}. It was found that the positive cosmological constant enhances the effect of dispersion and makes the black hole formation more difficult compared with the corresponding asymptotically flat spacetime. Recently with the growing interest in anti-de Sitter (AdS) spacetimes, a lot of works have been done on the gravitational collapse in AdS space \cite{Bizon_10,Garfinkle_11,Dias_11,Liebling_12,Wu_12,Maliborski_12,Buchel_12,Maliborski_13,Craps_14,Balasubramanian_14,Cai_15,Cai_16,Deppe_14}. The AdS spacetime is in stark contrast to asymptotically flat and de Sitter spacetimes because of the AdS boundary which needs to input suitable boundary conditions in order to make the evolution of the dynamical system well defined.  It was claimed that in AdS space black hole can always
be formed under arbitrarily small generic perturbations due to the
interplay of local nonlinear dynamics and the non-local kinematic
effect of the AdS reflecting boundary \cite{Bizon_10}. While in Einstein-Gauss-Bonnet AdS gravity, it was found that black holes will not form dynamically if the total mass/energy content of the spacetime is too small \cite{Deppe_14}. Further investigations of gravitational collapse in AdS spacetime are still being carried out.

Most phenomenologies of gravitational collapse were disclosed in Einstein gravity. It is of great interest to ask questions on how about the stability of the spacetime in modified gravity, whether black holes can be formed through gravitational collapse in modified gravity and whether there is a black hole threshold in the space of initial data in modified gravity.  The attempts to study the collapse of a free scalar field in the Brans-Dicke model of gravity were carried out in \cite{Liebling_96, Chiba_96}. It was found that  at the critical point of black hole formation, the model admits two distinctive solutions depending on the value of the coupling parameter between the Brans-Dicke field and matter, which admits one solution to be discretely self-similar and the other to exhibit continuous self-similarity. Recently the study of gravitational collapse has been extended to a more general gravity theory, the Horndeski theory \cite{Koutsoumbas_15}. However to avoid the complexity, the authors concentrated their attention on a homogeneous time-dependent scalar field coupling to curvature and Einstein tensor. In this paper, we are going to examine the gravitational collapse in a specific $f(R)$ gravity model, the Starobinsky $R^2$ gravity.  The evidences of existing spherically symmetric black holes in $f(R)$ gravity were reported for example in \cite{Canate_15}.  However the dynamical process of forming black holes in $f(R)$ gravity  through gravitational collapse is not clear. In the cosmological context, the $f(R)$ gravity can accommodate the early \cite{Starobinsky_80} or late time \cite{Carroll_04} acceleration of the universe. This implies that $f(R)$ gravity can produce some kind of repulsive force similar to that of inflaton or dark energy in the Einstein gravity. Whether such kind of repulsive effect in $f(R)$ gravity can hinder the gravitational collapse and the formation of black holes is a question to be answered. Thus we have strong motivation to generalize the formalism of studying the gravitational collapse well established in general relativity to the $f(R)$ gravity to examine the dynamical competitions in the gravitational collapse, disclose the threshold of black hole formation and compare with corresponding properties in the Einstein gravity.

The studies on gravitational collapse in $f(R)$ gravity have been reported from some aspects.  Fluid gravitational collapse in $f(R)$ gravity  was examined in  \cite{Sharif_10,Sharif_13,Sharif_14,Kausar_14,Chakrabarti_16}.  The gravitational collapse of a charged black hole in $f(R)$ gravity was studied  in \cite{Hwang_11}  by using  double-null formalism and the mass inflation of the Cauchy horizon was examined.  In \cite{Cembranos_12}, the authors analyzed a general $f(R)$ model with uniformly collapsing cloud of self-gravitating dust  particles.  Dark matter halo formation was studied in \cite{Kopp_13}.  Spherical scalar collapse in  $f(R)$ gravity towards a black hole formation was simulated in \cite{Guo_13}.  The  gravitational collapse of massive stars in $f(R)$ gravity was analyzed in  \cite{Goswami_14}.  Furthermore employing the double-null formalism, spherical scalar collapse for the Starobinsky $R^2$ model was investigated to explore the black hole and singularity formation scenarios in \cite{Guo_15}. It was argued that when matter field is strong enough, a black hole including a central singularity can be formed.
In this paper, we will apply the formalism developed by Choptuik to examine the evolution of a dynamical system obeying the gravity-matter equations in $f(R)$ gravity. We will disclose the threshold of black hole formation in the space of initial data for $f(R)$ gravity and examine carefully the nature of dynamical competition to explain where can the energy in the system of $f(R)$ gravity end up at late times. This can help us understand deeply on the gravitational collapse in $f(R)$ gravity from the point of view of perturbation theory and compare the properties with general relativistic collapsing models in Einstein gravity.

In our study, we will concentrate on the  asymptotically flat spacetime with spherical symmetry. We will examine the perturbation of massless scalar matter field $\psi$ minimally coupled to gravity. The organization of our paper is as follows:
In section \ref{sec:2}, we will write down the equations of motion for gravitational collapse. Then in section \ref{sec:3} we will explain numerical method. We will report numerical results in section \ref{sec:4}. In the final section we will present conclusions and discussions.

\section{\label{sec:2}Equations of motion for gravitational collapse in $f(R)$
	gravity}

We start with the four-dimension action of $f(R)$ gravity including a non-linear function $f$ in terms of Ricci scalar $R$ in the Jordan frame \cite{Sotiriou_08,Felice_10}
\begin{equation}
S_{J}=\int d^{4}x\sqrt{-g}\frac{1}{2\kappa^{2}}f(R)+\int d^{4}x\mathcal{L}_{M}(g_{\mu\nu},\psi),
\end{equation}
where $\kappa^{2}=8\pi G$, $g$ is the determinant of the metric
$g_{\mu\nu}$, $\mathcal{L}_{M}$ is the matter Lagrangian that depends
on $g_{\mu\nu}$ and matter  field $\psi$. For the Einstein gravity without cosmological constant,  $f(R)=R$. We consider the neutral
scalar field as matter field
\begin{equation}
\mathcal{L}_{M}(g_{\mu\nu},\psi)=-\frac{1}{2}\sqrt{g}g^{\mu\nu}\partial_{\mu}\psi\partial_{\nu}\psi.
\end{equation}

Varying the action  with respect to $g_{\mu\nu}$, we can obtain the gravity equation in the Jordan frame.
\begin{equation}
F R_{\mu\nu}-\frac{1}{2}f g_{\mu\nu}-{\nabla}_{\mu}{\nabla}_{\nu}F+g_{\mu\nu}{\nabla}^{\mu}{\nabla}_{\nu}F={\kappa}^2{T_{\mu\nu}^{M}}.
\end{equation}
Here $F(R)=\frac{df(R)}{dR}$ and ${T_{\mu\nu}^{M}}$ is the energy-momentum tensor of the matter field in the Jordan frame.
This is a fourth order differential equation for the  metric $g_{\mu\nu}$. To solve the equation, we have to impose boundary conditions up to the third order. However, there is not a natural choice for the high order boundary condition, which makes it difficult to solve the problem directly in the Jordan frame.

 It is more convenient to take a conformal transformation
\begin{equation}
\tilde{g}_{\mu\nu}=\Omega^{2}g_{\mu\nu},
\end{equation}
where $\Omega^{2}$ is the conformal factor.  In the literature, $\tilde{g}_{\mu\nu}$
is  referred to the metric in the Einstein frame while $g_{\mu\nu}$ represents the Jordan frame.  For the convenience of numerical calculation below, we will concentrate our investigation of the gravitational collapse of $f(R)$ gravity on the Einstein frame.

After the conformal transformation, we can rewrite the action in the Einstein frame as
\begin{equation}
S_{E}=\int d^{4}x\sqrt{-\tilde{g}}\left(\frac{1}{2\kappa^{2}}\tilde{R}-\frac{1}{2}\tilde{g}^{\mu\nu}\tilde{\nabla}_{\mu}\phi\tilde{\nabla}_{\nu}\phi-V(\phi)\right)+\int d^{4}x\mathcal{L}_{M}(F^{-1}(\phi)\tilde{g}_{\mu\nu},\psi), \label{eq:ActionEinstein}
\end{equation}
where $\phi$ is a new scalar field brought in by the conformal transformation, satisfying the relation
\begin{equation}
F(R)=\Omega^{2}=\exp\left(\sqrt{\frac{2}{3}}\kappa\phi\right).\label{eq:ConformalFactor}
\end{equation}
The scalar potential $V(\phi)=\frac{RF-f}{2\kappa^{2}F^{2}}.$ The
tilde covariant derivative $\tilde{\nabla}$ is compatible with $\tilde{g}_{\mu\nu}$.
$\tilde{R}$ is the Ricci scalar in the Einstein frame. From the action in  the Einstein frame, it is clear that the new scalar field $\phi$ is directly coupled to matter field $\psi$.
For Einstein gravity, $F(R)=\partial f/\partial R=1$ and $\phi=0$.
We will see  in the following that the equation for metric  $\tilde{g}_{\mu\nu}$ is of the second order and the natural boundary conditions can be imposed to solve these equations.

The Lagrangian density of  matter field $\psi$ in the Einstein frame has the form
\begin{equation}
\mathcal{L}_{M}=-\frac{1}{2}\sqrt{-\tilde{g}}\tilde{g}^{\mu\nu}F^{-1}\partial_{\mu}\psi\partial_{\nu}\psi.
\end{equation}
The Lagrangian density of the new scalar field $\phi$ is given by
\begin{equation}
\mathcal{L}_{\phi}=\sqrt{-\tilde{g}}\left(\frac{1}{2}\tilde{g}^{\mu\nu}\partial_{\mu}\phi\partial_{\nu}\phi+V(\phi)\right).
\end{equation}
From the action in  the Einstein frame, we can derive the Einstein equation
\begin{eqnarray}
\tilde{G}_{\mu\nu} & = & 8\pi\left(\tilde{T}_{\mu\nu}^{\phi}+\tilde{T}_{\mu\nu}^{M}\right),
\end{eqnarray}
where the energy-momentum tensor of matter field $\psi$ is
\begin{eqnarray}
\tilde{T}_{\mu\nu}^{M} & = & \frac{\partial_{\mu}\psi\partial_{\nu}\psi-\frac{1}{2}\tilde{g}_{\mu\nu}\tilde{g}^{\alpha\beta}\partial_{\alpha}\psi\partial_{\beta}\psi}{\exp\left(\sqrt{\frac{2}{3}}\kappa\phi\right)},
\end{eqnarray}
and the energy momentum tensor of the new scalar field $\phi$ reads
\begin{eqnarray}
\tilde{T}_{\mu\nu}^{\phi}=-\frac{2}{\sqrt{-\tilde{g}}}\frac{\delta\mathcal{L}_{\phi}}{\delta\tilde{g}^{\mu\nu}} & = & \partial_{\mu}\phi\partial_{\nu}\phi-\tilde{g}_{\mu\nu}\left(\frac{1}{2}\tilde{g}^{\alpha\beta}\partial_{\alpha}\phi\partial_{\beta}\phi+V(\phi)\right).
\end{eqnarray}

In order to see the coupling between the new scalar field $\phi$ and the matter field $\psi$ more explicitly, we take the variation of the action and derive the equation of motion of the matter field $\psi$ and the new scalar field $\phi$, respectively
\begin{eqnarray}
\tilde{\nabla}_{\mu}\tilde{\nabla}^{\mu}\psi & = & \sqrt{\frac{2}{3}}\kappa\tilde{g}^{\mu\nu}\partial_{\mu}\phi\partial_{\nu}\psi,\label{eq:Eq_matter}\\
\tilde{g}^{\mu\nu}\tilde{\nabla}_{\mu}\tilde{\nabla}_{\nu}\phi & = & \frac{\partial V(\phi)}{\partial\phi}-\frac{\kappa}{\sqrt{6}}\tilde{T}^{M},\label{eq:Eq_phi}
\end{eqnarray}
where $\tilde{T}^{M}=-\exp\left(-\sqrt{\frac{2}{3}}\kappa\phi\right)\tilde{g}^{\alpha\beta}\partial_{\alpha}\psi\partial_{\beta}\psi$
is the trace of the energy momentum tensor $\tilde{T}_{\mu\nu}^{M}$
of matter field $\psi$. It is obvious that the new scalar field $\phi$ is non-minimally coupled to the matter
field $\psi$ in the equations of motion.

In the cosmological context, the new scalar field $\phi$ appears in the Einstein frame can be interpreted as inflaton or dark energy fluid which accounts for the accelerated expansion of the universe. The new scalar degree of freedom introduced in the $R^2$ gravity can also account for the dark matter of our Universe \cite{Cembranos:2008gj,Cembranos:2010qd}. It is clear that the new scalar field $\phi$ is directly coupled to matter field $\psi$. More discussions on the interaction between dark energy and matter in cosmology can be found in a recent review \cite{Wang_16} and references therein.
 In this work we will concentrate on the gravitational collapse in $f(R)$ gravity in the Einstein frame. At the first glance, comparing to that of the Einstein gravity, the appearance of the new scalar field $\phi$ in the Einstein frame of $f(R)$ gravity playing the role of the repulsive effect in cosmology, which  can hinder the gravitational collapse and make the formation of black hole more difficult.   Whether our physical intuition is correct, we need to do careful investigations to check.

In numerical calculations, we have to specify the exact form of $f(R)$ gravity. Hereafter we take the ansatz
\begin{eqnarray}
f(R) & = & R+\alpha R^{n},
\end{eqnarray}
where $\alpha$ is a constant. From (\ref{eq:ConformalFactor}),
we can derive
\begin{eqnarray}
R & = & \left(\frac{1}{\alpha n}\left[\exp\left(\sqrt{\frac{2}{3}}\kappa\phi\right)-1\right]\right)^{\frac{1}{n-1}},\label{eq:JordanRicci}\\
V(\phi) & = & \frac{\alpha(n-1)\left(\frac{1}{\alpha n}\left[\exp\left(\sqrt{\frac{2}{3}}\kappa\phi\right)-1\right]\right)^{\frac{n}{n-1}}}{2\kappa^{2}}\exp\left(-2\sqrt{\frac{2}{3}}\kappa\phi\right).\label{eq:ScalarPotential}
\end{eqnarray}
Now if we take the initial condition of $\phi$ as a wave tends to
$0$ when the spatial radius $r\rightarrow\infty$, the scalar potential
or the Ricci scalar in the Jordan frame will diverge in asymptotic flat
spacetime as $r\rightarrow\infty$ if $n<1$. So we have to concentrate on
$n>1$. (The case $n=1$ is just the general relativity). However,
in numerical calculations, we found that there will be convergence
problem due to radical sign in (\ref{eq:JordanRicci},\ref{eq:ScalarPotential})
when $n\neq2$. So we fix $n=2$ through out this paper. The model
$f(R)=R+\alpha R^{2}$ is often called Starobinsky $R^{2}$ model
which describes the inflation in the early universe. For details of the model, see
\cite{Starobinsky_80} and the review \cite{Felice_10,Vilenkin_85}.

Now we start to derive the field equations governing the gravitational collapse in spherical symmetry by using the Choptuik formalism. We choose the spherically symmetric background in four-dimensional
asymptotically flat spacetime with the metric
\begin{eqnarray}
ds^{2} & = & -A(t,r)e^{-2\delta(t,r)}dt^{2}+\frac{1}{A(t,r)}dr^{2}+r^{2}(d\theta^{2}+\sin^{2}\theta d\varphi^{2}),
\end{eqnarray}
where $A(t,r)$ and $\delta(t,r)$ are functions of time $t$ and
spatial coordinate $r$. $r$ measures the proper surface area in this
coordinate system.

Introducing auxiliary variables for the new scalar field $\phi$ and matter field $\psi$,
\begin{eqnarray}
Q(t,r)=\partial_{r}\phi(t,r) & , & P(t,r)=A^{-1}e^{\delta}\partial_{t}\phi(t,r),\\
\Phi(t,r)=\partial_{r}\psi(t,r) & , & \Pi(t,r)=A^{-1}e^{\delta}\partial_{t}\psi(t,r),\nonumber
\end{eqnarray}
we can rewrite equations of motion (\ref{eq:Eq_matter},\ref{eq:Eq_phi}) into
\begin{eqnarray}
\partial_{t}Q & = & \partial_{r}\left(Ae^{-\delta}P\right),\nonumber \\
\partial_{t}P & = & \frac{1}{r^{2}}\partial_{r}\left(r^{2}Ae^{-\delta}Q\right)-e^{-\delta}\frac{\partial V(\phi)}{\partial\phi}+\frac{\kappa\exp\left(-\sqrt{\frac{2}{3}}\kappa\phi\right)}{\sqrt{6}}Ae^{-\delta}\left(\Pi^{2}-\Phi^{2}\right),\nonumber \\
\partial_{t}\Phi & = & \partial_{r}\left(Ae^{-\delta}\Pi\right),\label{eq:evolutionScalar}\\
\partial_{t}\Pi & = & \frac{1}{r^{2}}\partial_{r}\left(r^{2}Ae^{-\delta}\Phi\right)+\sqrt{\frac{2}{3}}\kappa Ae^{-\delta}\left(\Pi P-\Phi Q\right).\nonumber
\end{eqnarray}
The nontrivial Einstein equations in spherical symmetric spacetime
are
\begin{eqnarray}
\partial_{r}A & = & \frac{1-A}{r}-4\pi rA\left[P^{2}+Q^{2}+\left(\Pi^{2}+\Phi^{2}\right)\exp\left(-\sqrt{\frac{2}{3}}\kappa\phi\right)\right]-8\pi rV(\phi),\label{eq:metricA}\\
\partial_{r}\delta & = & -4\pi r\left[P^{2}+Q^{2}+\left(\Pi^{2}+\Phi^{2}\right)\exp\left(-\sqrt{\frac{2}{3}}\kappa\phi\right)\right].\label{eq:metricDelta}
\end{eqnarray}
In addition, there is also a momentum constraint equation
\begin{equation}
\partial_{t}A=-8\pi rA^{2}e^{-\delta}\left[PQ+\Pi\Phi\exp\left(-\sqrt{\frac{2}{3}}\kappa\phi\right)\right].\label{eq:MomentumConstraint}
\end{equation}

Taking $f(R)=R+R^{2}$, the scalar potential
\begin{eqnarray}
V(\phi) & = & \frac{\left[\exp\left(\sqrt{\frac{2}{3}}\kappa\phi\right)-1\right]^{2}}{8\kappa^{2}}\exp\left(-2\sqrt{\frac{2}{3}}\kappa\phi\right).
\end{eqnarray}
The Ricci scalar in the Einstein frame can be calculated from Einstein
equations
\begin{equation}
\tilde{R}=8\pi\left(\left[\left(Q^{2}-P^{2}\right)+\frac{1}{\exp(\sqrt{\frac{2}{3}}\kappa\phi)}\left(\Phi^{2}-\Pi^{2}\right)\right]A+4V(\phi)\right).
\end{equation}
Finally, the gravitational potential is defined in the form
\begin{equation}
V(t,r)=\frac{1-g{}_{00}(t,r)}{2}=\frac{1}{2}\left(1+A(t,r)e^{-2\delta(t,r)}\right).
\end{equation}
The gravitational potential can give us clear picture to understand whether the mass/energy can be trapped and black
holes can be formed or not.

\section{\label{sec:3}Numerical method}

We have derived all the equations needed to simulate the gravitational
collapse in $f(R)$ gravity in the last section. Now we need to specify
initial and boundary conditions on how to solve these equations.

The free data for the system are the auxiliary variables at initial moment $t=0$. We choose simple initial conditions of this system
\begin{eqnarray}
\phi(0,r)=a_{1}\exp\left[-\left(\frac{r}{\sigma_{1}}\right)^{2}\right] & , & \psi(0,r)=a_{2}\exp\left[-\left(\frac{r}{\sigma_{2}}\right)^{2}\right],\label{eq:initialCondition}\\
P(0,r)=Q(0,r)=\partial_{r}\phi(0,r) & , & \Pi(0,r)=\Phi(0,r)=\partial_{r}\psi(0,r),\nonumber
\end{eqnarray}
which means that the initial new scalar field $\phi$ and matter field $\psi$ are ingoing waves. We
fix $\sigma_{1}=\sigma_{2}=4$ in the numerical computation. $a_1$ and $a_2$ are two amplitudes of the initial perturbations which will be tuned to find the threshold of black hole formation.

Besides the initial conditions, we also need to specify the boundary
conditions. The regularity of (\ref{eq:metricA}) at center $r=0$
requires that $A(t,0)=0$. Solving the constraint equation (\ref{eq:metricDelta})
by integrating outwards we need to take the boundary condition $\delta(t,0)=0$.
This implies that the time coordinate at center $r=0$ is chosen as
its proper time.

Given initial conditions (\ref{eq:initialCondition}), the initial
metric can be derived from (\ref{eq:metricA},\ref{eq:metricDelta}).
The time evolutions of the new scalar field $\phi$ and the matter field $\psi$ can then be determined by (\ref{eq:evolutionScalar}).
After we get the evolutions of two fields at the next moment, the corresponding new metric
can be worked out from (\ref{eq:metricA},\ref{eq:metricDelta}).
Repeating this procedure, we can obtain the evolutions of this system.
Though the constraint equation (\ref{eq:MomentumConstraint}) is not
used in obtaining the system evolution, it can be used to check the accuracy of the
numerical computations at each step.

In the numerical calculation, the criteria for the black hole formation is set to be $A(T_{B},r_{B})<0.05$ where $r_{B}$ is the apparent horizon
of the black hole, $T_{B}$ is the time needed to form the apparent horizon. The result of the Ricci scalar at the center when the criteria is satisfied to form the black hole is instructive and will be shown in the following.

We solve the system numerically using fourth-order Runge-Kutta method
in both time and spatial directions. The adjustable time step $\Delta t$
is kept $1/6<e^{-\delta(\xi)}\Delta t/\Delta r<1/3$ for the spatial
grid spacing $\Delta r$. Here $\xi$ is the cut off of the spatial radius
used in numerical calculation. This scheme allows for stable long-time
evolution. To overcome the error caused by $1/r$ in the evolution
equations near $r=0$, we expand the variables around the center.
Smoothness at the origin implies that we have following power series
expansions near $r=0$:
\begin{eqnarray}
A(t,r) & = & 1+A_{2}(t)r^{2}+A_{4}(t)r^{4}+A_{6}(t)r^{6}+O(r^{7}),\nonumber \\
\delta(t,r) & = & \delta_{2}(t)r^{2}+\delta_{4}(t)r^{4}+\delta_{6}(t)r^{6}+O(r^{7}),\nonumber \\
\phi(t,r) & = & \phi_{0}(t)+\phi_{2}(t)r^{2}+\phi_{4}(t)r^{4}+O(r^{5}),\\
\psi(t,r) & = & \psi_{0}(t)+\psi_{2}(t)r^{2}+\psi_{4}(t)r^{4}+O(r^{5}),\nonumber
\end{eqnarray}
in which we have considered $A(0,r)=1$ and $\delta(0,r)=0$. These
expansions allow us to express the function at the boundary with its
values inside the domain. For example, assuming expansion $f(r)=f_{0}+f_{2}r^{2}+f_{4}r^{4}+f_{5}r^{6}+O(r^{7})$
and taking a small interval $\triangle r$, we have
\begin{eqnarray}
f(\triangle r) & = & f_{0}+f_{2}(\triangle r)^{2}+f_{4}(\triangle r)^{4}+f_{6}(\triangle r)^{6},\nonumber \\
f(2\triangle r) & = & f_{0}+f_{2}(2\triangle r)^{2}+f_{4}(2\triangle r)^{4}+f_{6}(2\triangle r)^{6},\\
f(3\triangle r) & = & f_{0}+f_{2}(3\triangle r)^{2}+f_{4}(3\triangle r)^{4}+f_{6}(3\triangle r)^{6},\nonumber \\
f(4\triangle r) & = & f_{0}+f_{2}(4\triangle r)^{2}+f_{4}(4\triangle r)^{4}+f_{6}(4\triangle r)^{6}.\nonumber
\end{eqnarray}
Then $f_{0}$ can be written in terms of $f(\triangle r),f(2\triangle r),f(3\triangle r)$
and $f(4\triangle r)$ as
\begin{eqnarray}
f(0)=f_{0} & = & \frac{56f(\triangle r)-28f(2\triangle r)+8f(3\triangle r)-f(4\triangle r)}{35}.
\end{eqnarray}
Similarly, we can also write $f(\triangle r)$ in terms of $f(2\triangle r),f(3\triangle r),f(4\triangle r),f(5\triangle r)$.
These expansions allow us to express the function at the boundary
with its values inside the domain.

Besides, de l'H\"{o}spital rule $f/r=f'(r)-r(f/r)'$ is applied at $r=0$
if $f(r)\rightarrow0$ as $r\rightarrow0$. The instability of $f/r$
near $r=0$ is suppressed by the factor $r$ comparing to the leading
term $f'(r)$ \cite{Maliborski_13_2}. The Kreiss-Oliger dissipation
which is crucial to stabilize the solutions is employed \cite{Kreiss_72}.

\section{\label{sec:4}Numerical results }
\begin{figure}[h]
	\begin{centering}
		\includegraphics[scale=0.85]{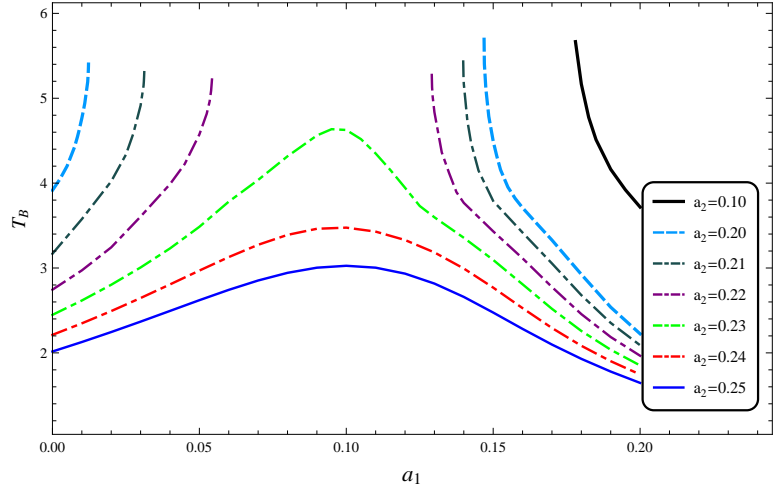}
		\par\end{centering}
	
	\caption{\label{fig:FormationTime}{\small{}The time needed to form the apparent horizon
			of black hole with respect to different $a_{1}$. The lines from up
			to down denote $a_{2}=0.1,0.2,0.21,0.22,0.23,0.24,0.25$, respectively.}}
\end{figure}

Now we report the phenomena of gravitational collapse in $f(R)$ gravity. We fix  $\alpha=1$ and $n=2$ in our numerical calculations. We exhibit the time scale for the black hole formation and its relation to the amplitudes of initial perturbations of the new scalar field and matter field perturbations in Fig.\ref{fig:FormationTime}.

Different colors of lines indicate different strengths of the perturbation of the matter field $\psi$.  We see that when the initial amplitude of the matter field perturbation is stronger, the time scale for the formation of the apparent horizon is shorter which means that the black hole can be formed more easily. This result is consistent with that observed in the Einstein gravity \cite{Gundlach_07}.

From the definition of the new scalar field $\phi$, we know that its strength indicates the deviation of the $f(R)$ gravity from the Einstein gravity. In the cosmological context, this new scalar field plays the role of inflaton or dark energy, acting as a repulsive force accounts for the accelerated expansion of the universe. When we fix the amplitude of the matter field perturbation $a_2$, we see that with the increase of the amplitude of the new scalar perturbation $a_1$, the repulsive effect brought in by the new scalar field dilutes the matter field perturbation and hinders the formation of the black hole so that it needs longer period of time for the apparent horizon to be developed.  When the matter field perturbation is not strong enough, the increase of $a_1$ can even prevent the formation of the black hole through gravitational collapse. However when the amplitude of the new scalar perturbation $a_1$ is strong enough, we have observed some interesting results contrary to our naive understanding inherited from cosmology. Instead of prevention, the strong enough perturbation of the new scalar field can even stimulate the formation of black hole from gravitational collapse. The strong perturbation of the new scalar field (similar to the dark energy field) in comparable to the perturbation of the matter field does not exist in the cosmological scale. But it was discussed that the inhomogeneous perturbation of dark energy, although it is small, can contribute to the structure formation in cosmology \cite{Wang_10}\cite{aa}. In the small scale, here we have seen that when the perturbation of the new scalar field is strong enough and is comparable to that of the matter field, it not only participates, but even stimulates the structure formation and makes the gravitational collapse quicker to form the black hole.

Dynamically, in addition to the kinetic energy of the matter field $\psi$ and the gravitational potential, the new scalar field $\phi$ adds a new competition force in the system. This is very similar to the study of the gravitational collapse of charged scalar field in our previous work \cite{Zhang_15}. We have seen that the dynamical competition can be controlled by tuning the parameters $a_1$ and $a_2$ in the initial perturbations of the new scalar field $\phi$ and the matter field $\psi$.

We want to emphasis that the phenomena we discovered are independent
of the special types of initial conditions (\ref{eq:initialCondition}) we have chosen.
We have checked that the numerical results hold with other kinds of initial
conditions such as
\begin{eqnarray}
\phi(0,r)=a_{1}\exp\left[-\left(\frac{r}{\sigma_{1}}\right)^{2}\right]r^{2} & , & \psi(0,r)=a_{2}\exp\left[-\left(\frac{r}{\sigma_{2}}\right)^{2}\right]r^{2},\nonumber \\
\phi(0,r)=a_{1}\cos\left(\frac{\pi}{2}\tanh(r)\right) & , & \psi(0,r)=a_{2}\cos\left(\frac{\pi}{2}\tanh(\frac{r}{2})\right),\\
\phi(0,r)=a_{1}\cos\left(\frac{\pi}{2}\tanh(r)\right)r^{2} & , & \psi(0,r)=a_{2}\cos\left(\frac{\pi}{2}\tanh(\frac{r}{2})\right)r^{2}.\nonumber
\end{eqnarray}
Similar qualitative results to Fig.\ref{fig:FormationTime} have been discovered
for different initial conditions. Thus the physical results  we have obtained are
universal.

In order to understand the phenomena in the above figure more clearly, we show the property of the gravitational potential in the following and examine how it evolves to trap the collapsing fields to form the black hole. Fixing the amplitude of the new scalar field perturbation $a_1$, the evolutions of the gravitational potential due to the change of the amplitude of matter field perturbation $a_2$ are shown in Fig.2.

\begin{figure}[!htbp]
	\centering
	\includegraphics[width=0.45\textwidth]{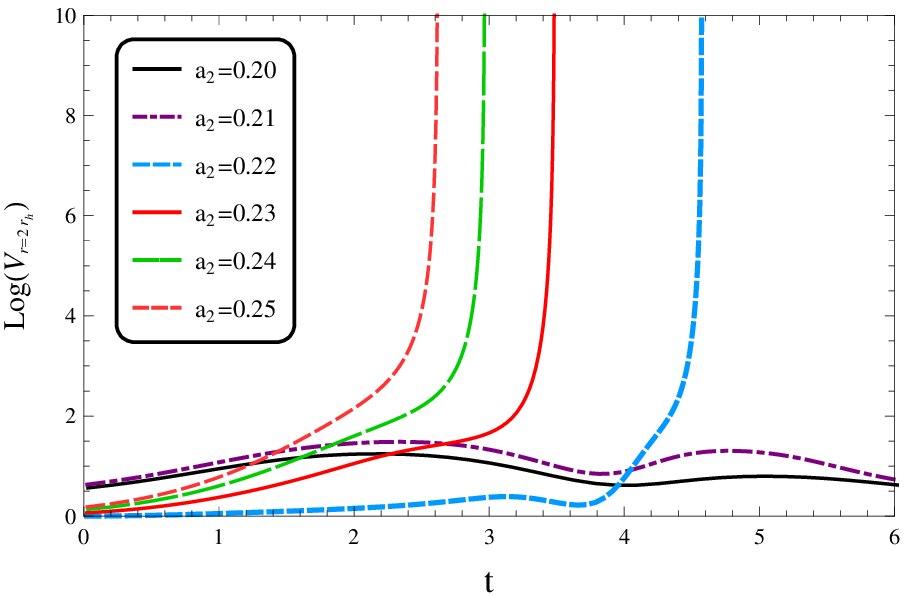}\qquad
	\includegraphics[width=0.45\textwidth]{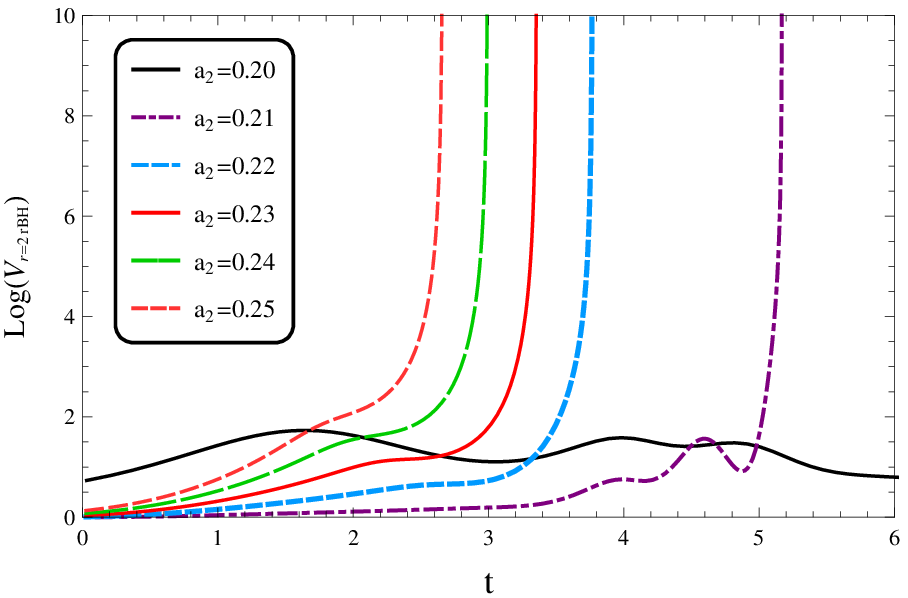}
	\caption{{\small{}\label{fig:Comparison0514}The evolution of the gravitational potential
			in the Einstein frame when $a_{1}=0.05$ (left) and $0.14$ (right), respectively.}}
\end{figure}

It is easy to see that when the initial amplitude of the matter field perturbation is stronger, the gravitational potential grows up quicker so that the collapsing field is easier and quicker to be trapped in a small region to form a black hole.

\begin{figure}[h]
	\begin{centering}
		\begin{tabular}{cc}
			\includegraphics[scale=0.8]{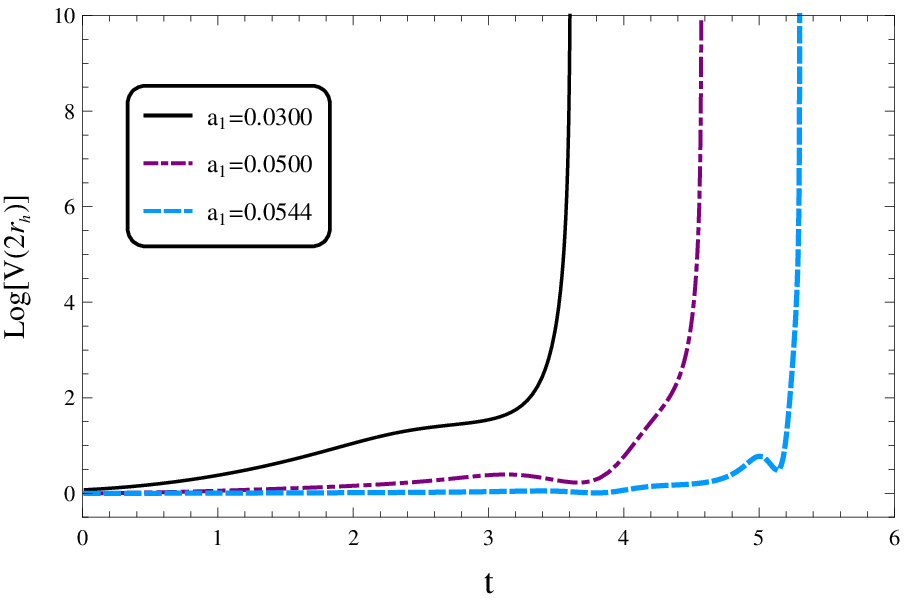} & \includegraphics[scale=0.8]{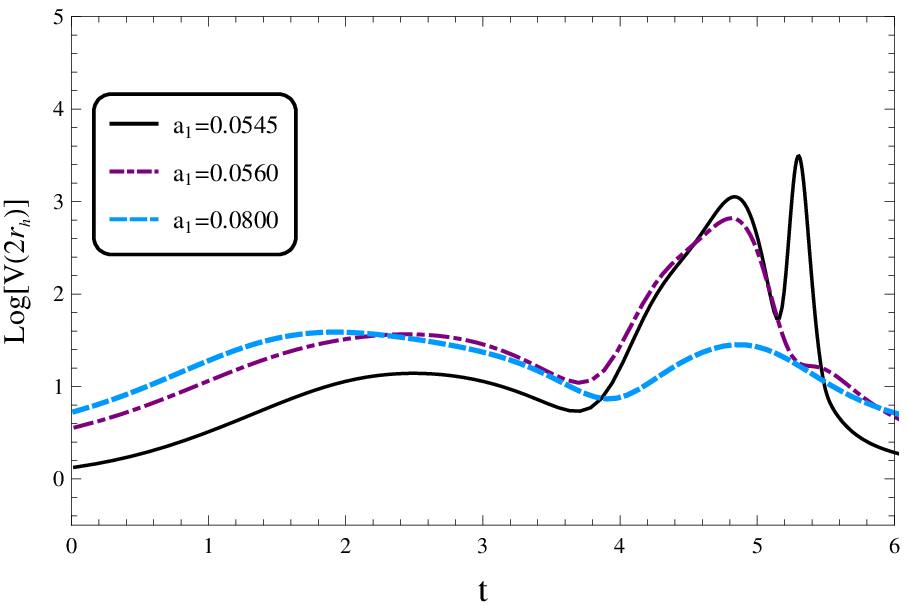}\tabularnewline
			\includegraphics[scale=0.8]{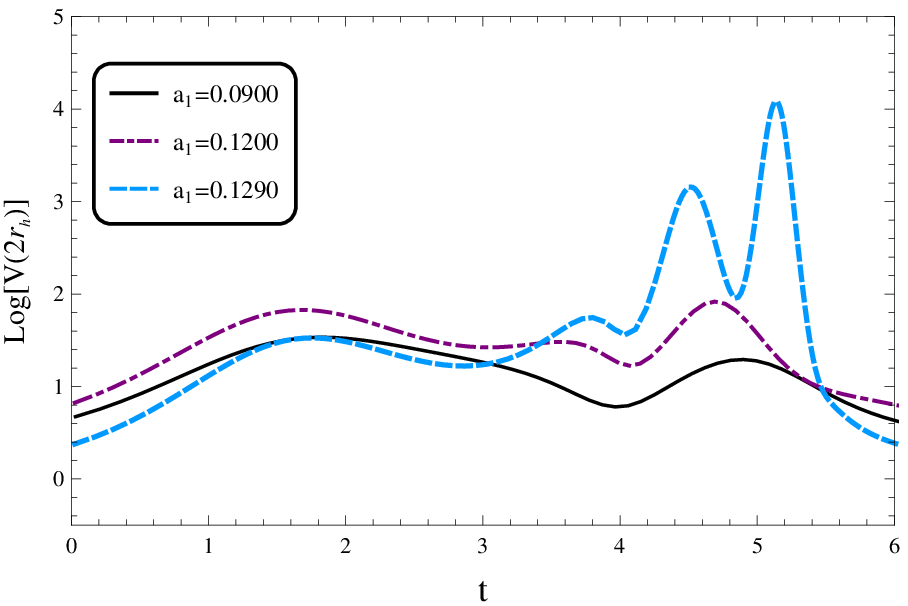} & \includegraphics[scale=0.8]{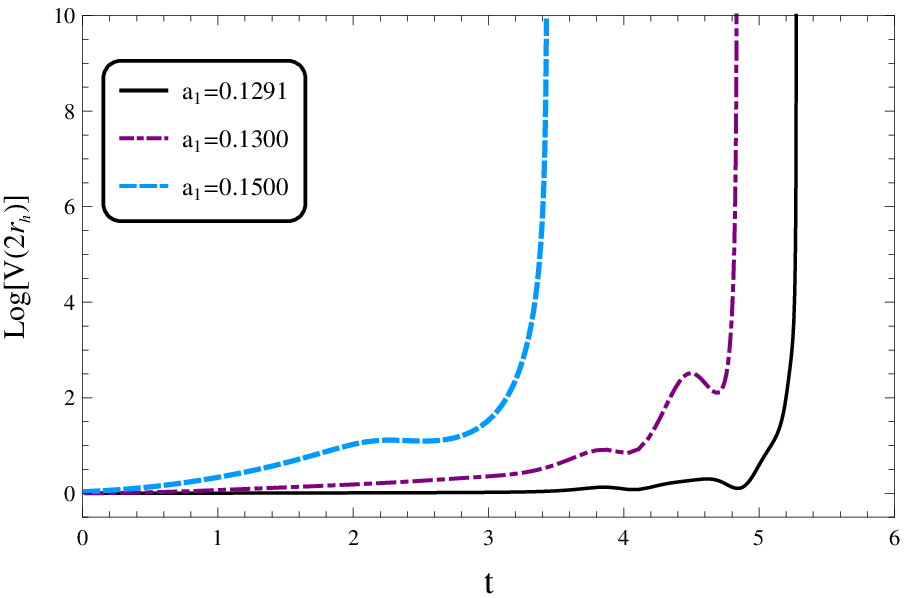}\tabularnewline
		\end{tabular}
		\par\end{centering}
	
	\caption{\label{fig:GravPotential}{\small{}The time evolution of the gravitational potential
			at $r=2r_{h}$. We fix $a_{2}=0.22$ and let $a_{1}$ vary. Here
			$r_{h}$ is the apparent horizon of the black hole. If there are not
			black hole, $r_{h}$ is the position where $a(t,r)$ is minimum when
			the iteration stops. }}
\end{figure}

In Fig.\ref{fig:GravPotential} we fix the initial amplitude of the matter perturbation $a_2$ and exhibit the dependence of the evolution of the gravitational potential on the initial amplitude of the new scalar field perturbation $a_1$.

In the upper left panel, we see that for the fixed amplitude of the matter field perturbation $a_2$, it takes longer time for the gravitational potential to grow when $a_1$ becomes bigger. This actually tells that more
bounces of the matter field in the process of collapse are needed before the black hole formation. For $a_{1}=0.03$
the matter field needs two bounces and for $a_{1}=0.0544$ it needs four
bounces to settle down to form black hole. In this panel the amplitude of the new scalar field  perturbation $\phi$ is minor, thus although it plays the repulsive role, it cannot prevent the formation of the gravitational barrier to trap the matter field.

The upper right panel shows the gravitational potential behavior when
$0.0545\leq a_{1}\leq0.08$. We see that the gravitational potential
is not strong enough to trap the matter field so that the matter field can run away instead of collapsing. The failure of forming a black hole is because of the strong repulsive force contributed by the new scalar field
$\phi$. As $a_{1}$ increases, the peak of the gravitational
potential becomes even smaller. The repulsion of the new scalar field $\phi$ now dominates
the competition of dynamics.

In the lower left panel we exhibit the gravitational potential behavior when
$0.09\leq a_{1}\leq0.129$. Here something interesting happens. Though
the repulsion of $\phi$ still wins the competition and gravitational
potential is not strong enough to trap the matter field, the peak of gravitational
potential becomes higher and higher as $a_{1}$ increases.

When $a_{1}>0.129$, the gravitational potentials are shown in the lower right panel.  The gravitational potential
becomes so big that black hole can be formed again. As $a_{1}$ increases, we see that
the peak of the gravitational potential appears earlier and fewer bounces
are needed for the matter field to settle down and being trapped to form the black hole. For example, there are four bounces of the matter field
before the formation of black hole when $a_{1}=0.1291$. However,
only two bounces are needed to form the black hole when $a_{1}=0.15$.
The contribution of the strong perturbation in the small scale increases the gravitational binding effect so that the black hole formation through collapse becomes easier.

To be more instructive, we plot the the evolution of the Ricci scalar at $r=0$ in the Einstein frame for fixing $a_1$ first in Fig.4.

\begin{figure}[!htbp]
	\centering
	\includegraphics[width=0.45\textwidth]{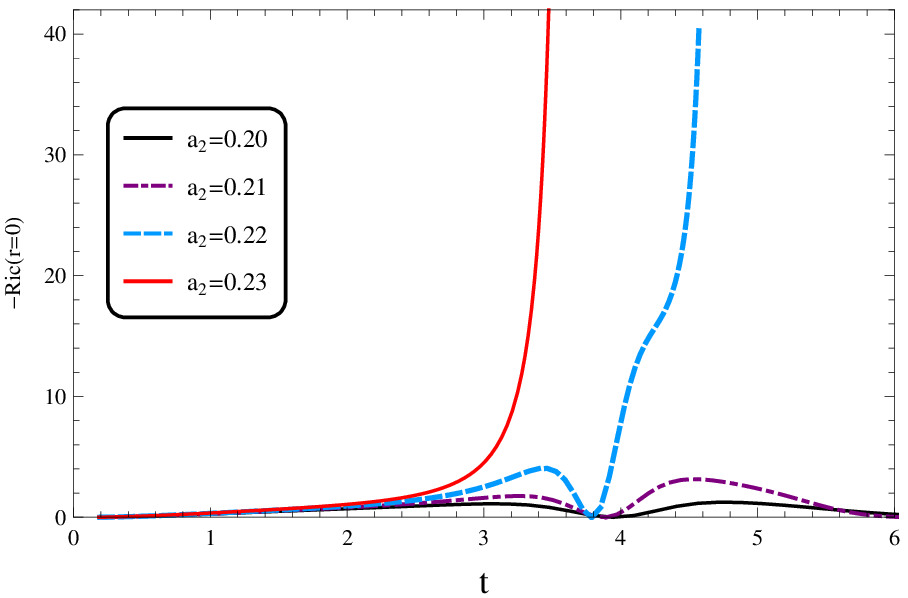}\qquad
	\includegraphics[width=0.45\textwidth]{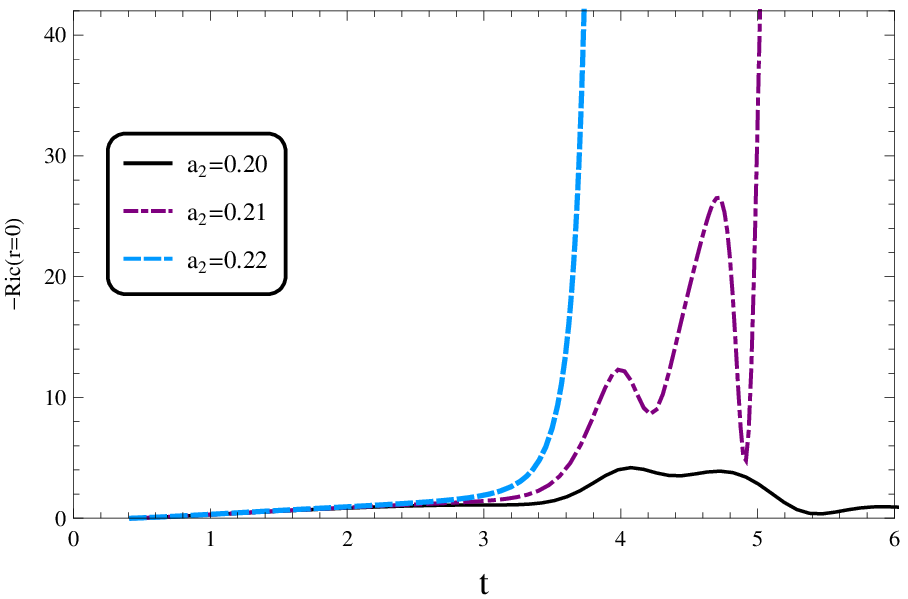}
\caption{\label{fig:Comparison0514}{\small{}The evolution of the Ricci scalar in the Einstein frame at center $r=0$
		when $a_{1}=0.05$ (left) and $0.14$ (right), respectively.}}
\end{figure}

We see that for strong initial amplitude of the matter field perturbation $a_2$, the Ricci scalar diverges quicker at the center which indicates the black hole is more easily to be formed from the collapse.

\begin{figure}[h]
	\begin{centering}
		\begin{tabular}{cc}
			\includegraphics[scale=0.8]{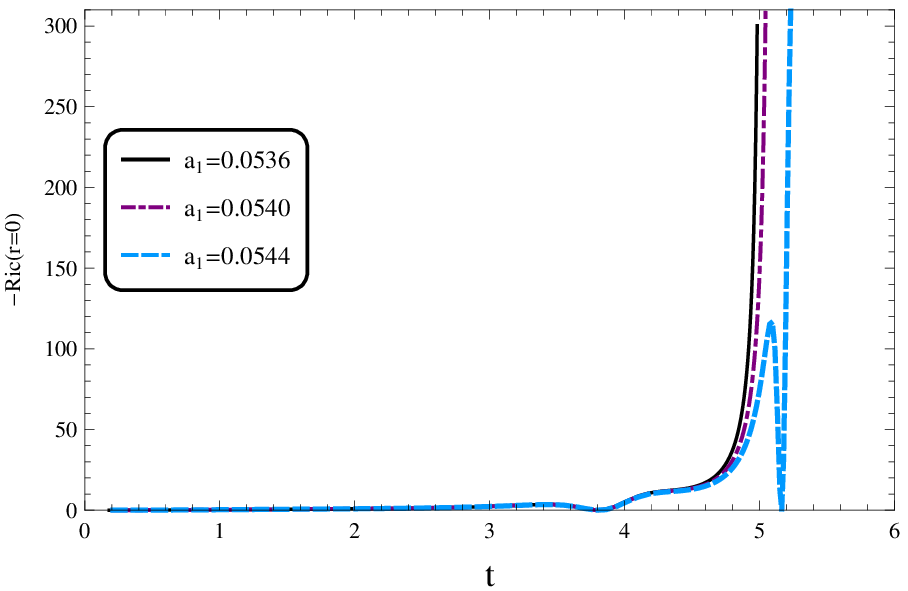} & \includegraphics[scale=0.8]{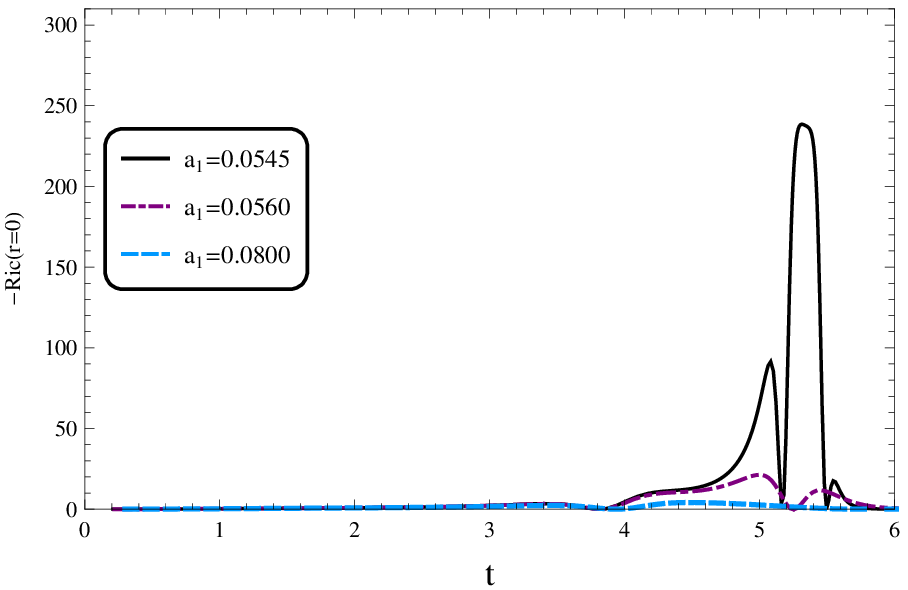}\tabularnewline
			\includegraphics[scale=0.8]{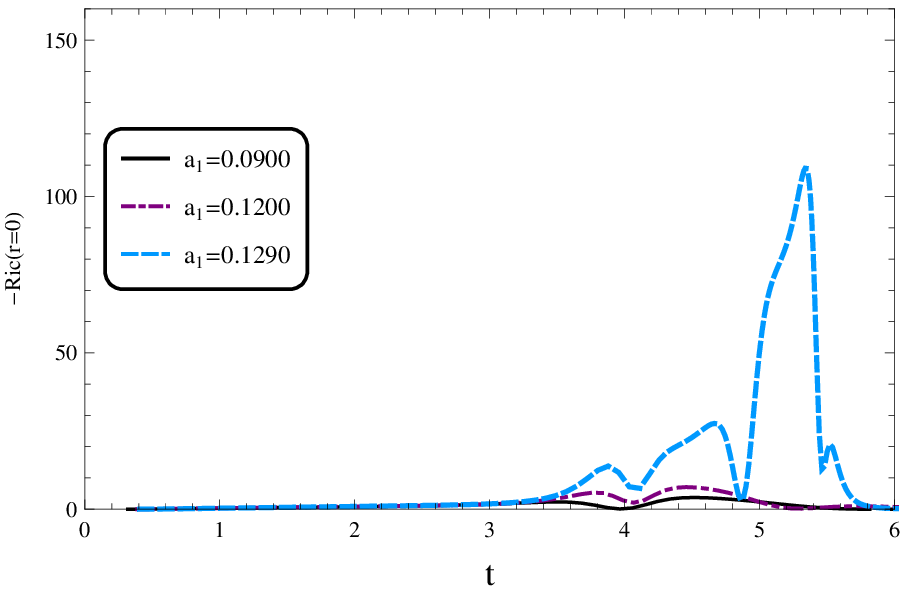} & \includegraphics[scale=0.8]{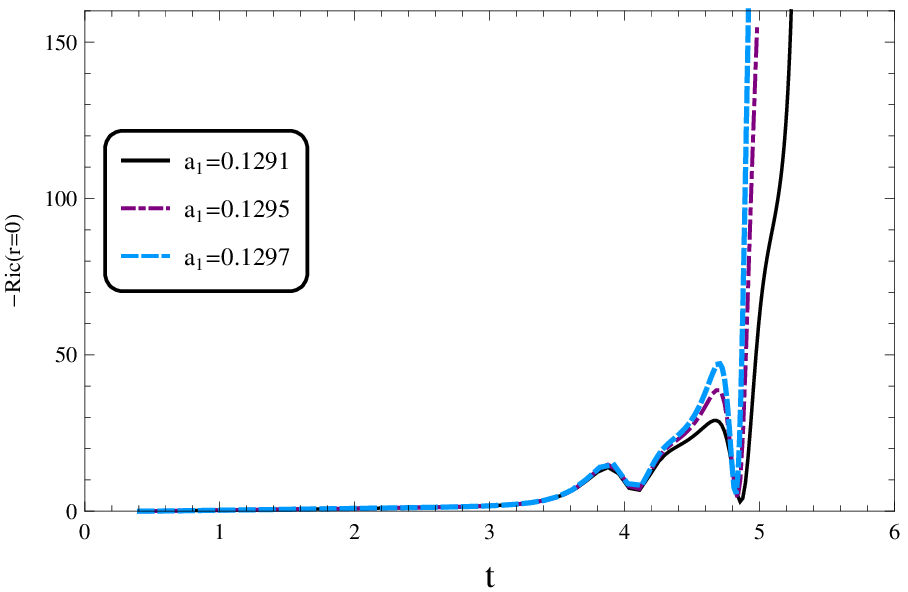}\tabularnewline
		\end{tabular}
		\par\end{centering}
	
\caption{\label{fig:Ricci}{\small{}The evolution of Ricci scalar in the Einstein frame
		at center $r=0$ with respect to various $a_{1}$. We have fixed  $a_{2}=0.22$.}}
\end{figure}

Fixing the initial amplitude of the matter field perturbation $a_{2}=0.22$, the dependences of the Ricci scalar evolution on the initial amplitude of the new scalar field perturbation  $a_{1}$ are shown in Fig.\ref{fig:Ricci}.

We see that at the center the Ricci scalar in the Einstein frame tends to diverge
when the black hole can be formed, while converge when the collapsing field is dispersed. The influence of the $a_1$ on the time scale for the blow up of the Ricci curvature is consistent with that in the gravitational potential in Fig.\ref{fig:GravPotential}. Actually this gives us further understanding on the different influences of the new scalar field and the matter field in the formation of black hole through the gravitational collapse.

\section{\label{sec:5}Conclusions and discussions}

In this paper we have studied the gravitational collapse in the model of a spherically symmetric, massless scalar field $\psi$, minimally coupled to gravity in a specific $R^2$ model in $f(R)$ gravity. Adopting the usual treatment, we have done the conformal transformation and expressed the $f(R)$ gravity in the Einstein frame. The conformal transformation from Jordan frame to Einstein frame brought a new scalar field to the system which is coupled to the matter field.
 In the cosmological context this new scalar field plays the role of the repulsive force just like the inflaton or dark energy in the early or late time accelerated expansion of the universe. On the other hand, this field can also work as dark matter \cite{Cembranos:2008gj,Cembranos:2010qd}. 
Thus at the first sight, the black hole formation through gravitational collapse in the $f(R)$ gravity contains more competitions between repulsive and attractive effects in dynamics than that in the Einstein general gravity.

Our numerical calculation focused on the Starobinsky $R^2$ gravity model, which can account for the inflation in the early universe. Adopting the Choptuik's formalism, we have studied the evolution of the dynamical system satisfying the matter-gravity equations. We have found that the formation of the black hole from the gravitational collapse depends on the thresholds of initial amplitudes of perturbations in the new scalar field introduced from conformal transformation and in the original matter field perturbation. Fixing the newly scalar field introduced by conformal transformation, we found that when the amplitude of the initial matter field perturbation is strong, the black hole can be formed more easily and more quickly. When we fix the amplitude of the initial matter perturbation field, with the increase of the amplitude of the initial scalar perturbation brought by the conformal transformation in the Einstein frame, the formation of the black hole from the gravitational collapse is hindered at first due to the repulsive effect of this new scalar field. If the matter perturbation is weak, the repulsive effect of the new scalar field can even prevent the formation of the black hole. However when the initial perturbation of the scalar field can be comparable to that of the matter field, when we increase further the initial amplitude of the scalar perturbation, we have observed that this new scalar perturbation can participate the structure formation in the gravitational collapse and stimulates the black hole formation. This result in the small scale gravitational collapse is different from our usual understanding from the cosmological context in the large scale.

We have examined carefully the competitions in dynamics by investigating the gravitational potential  in the system which plays the major role to trap the collapsing fields and contributes to the black hole formation.  It is interesting that we have clearly shown that the dynamical competition can be controlled by tuning the parameters in the initial perturbations in the new scalar field and the matter field. To show the result more instructive, we have also calculated the Ricci scalar at the center and its dependence of the tuning of parameters in the initial perturbations.
 These results can help us understand better the gravitational collapse in the $f(R)$ gravity.

We have transformed the results back to the Jordan frame from that we obtained in the Einstein frame. We see that the location of the horizon in the Jordan frame is the same as that in the Einstein frame since the conformal factor is regular at the horizon. Considering that the standard
 matter are minimally coupled to the metric in the Jordan Frame,  it would be very interesting to know how the collapse process looks in
the Jordan frame. But direct computation in the Jordan frame is difficult as we discussed, since we can not specify natural high order boundary conditions from physics to solve the fourth order differential equations. It was argued that Jordan frame and Einstein frame are equivalent, especially in the cosmological context \cite{He2012,Domenech2016,Chiba2013}, which implies that the properties of gravitational collapse we obtained in the Einstein frame also should hold in the Jordan frame. The relation between the Einstein frame and Jordan frame was also discussed in \cite{Sotiriou_08,Felice_10}. To disclose exactly how the gravitational collapse happens in the Jordan frame, careful investigation is still called for.

\section*{ACKNOWLEDGMENTS }

We thank Li-Wei Ji, Shao-Jun Zhang and Yun-Qi Liu for helpful discussions. B.W. acknowledges the discussion with E. Papantonopoulos.
This work was partially supported by NNSF of China.

\end{document}